\documentclass[conference,compsoc]{IEEEtran}

\IEEEoverridecommandlockouts

\usepackage[utf8]{inputenc}
\usepackage{graphicx}
\usepackage[dvipsnames]{xcolor}
\usepackage{colortbl}
\usepackage{epstopdf} 
\usepackage{verbatim}
\usepackage{pgfplotstable,filecontents}
\pgfplotsset{compat=1.9}
\usepackage{enumitem}
\usepackage{ulem}
\usepackage{breakcites}
\usepackage[colorlinks = true,
            linkcolor = BlueViolet,
            urlcolor  = BlueViolet,
            citecolor = BlueViolet,
            anchorcolor = BlueViolet]{hyperref}

\usepackage{soulutf8} 

\usepackage[numbers,sort&compress,square]{natbib} 
\usepackage{wrapfig}
\usepackage{longtable,tabu}

\usepackage{multirow}
\usepackage{multicol}
\usepackage{booktabs}
\usepackage{threeparttable}

\usepackage{caption}
\usepackage{subcaption}
\usepackage{paralist} 
\newenvironment{myitemize}
    { \begin{compactitem}[\leftmargin = 0pt $\circ$] }
    { \end{compactitem} }

\usepackage[colorinlistoftodos,prependcaption, textsize=small]{todonotes}

\usepackage{rotating}
\usepackage{lipsum}

\usepackage{amsmath, breqn}
\newcommand{\smalltilde}{\raise.17ex\hbox{$\scriptstyle\mathtt{\sim}$}} 

\usepackage{fancyhdr}

\AtBeginDocument{%
  \providecommand\BibTeX{{%
    \normalfont B\kern-0.5em{\scshape i\kern-0.25em b}\kern-0.8em\TeX}}}

\setdescription{topsep=0pt, partopsep=0pt, parsep=0pt, itemsep=0pt, leftmargin=.15cm}

\usepackage{changepage}

\begin{document}
\title{AI ATAC 1: An Evaluation of Prominent Commercial Malware Detectors
    \thanks{\scriptsize{
        This manuscript has been co-authored by UT-Battelle, LLC under Contract No. DE-AC05-00OR22725 with the US DOE. The United States Government retains and the publisher, by accepting the article for publication, acknowledges that the United States Government retains a non-exclusive, paid-up, irrevocable, world-wide license to publish or reproduce the published form of this manuscript, or allow others to do so, for United States Government purposes. The DOE will provide public access to these results of federally sponsored research in accordance with the DOE Public Access Plan (\url{http://energy.gov/downloads/doe-public-access-plan}).}}
        \thanks{\scriptsize{
        The research is based on work supported by the US Department of Defense (DOD), Naval Information Warfare Systems Command (NAVWAR), via the US Department of Energy (DOE) under contract  DE-AC05-00OR22725. The views and conclusions contained herein are those of the authors and should not be interpreted as representing the official policies or endorsements, either expressed or implied, of the DOD, NAVWAR, or the US Government. The US Government is authorized to reproduce and distribute reprints for governmental purposes notwithstanding any copyright annotation thereon. 
        }}
        \thanks{\scriptsize{
        MIT licensed code containing base classes for the software described in this paper, as well as an example implementation of the framework is provided \cite{framework2022code}.
        }}
        \thanks{\scriptsize{Authors thank Charlie Horak for the editorial review and Mike Karlbom for the guidance. 
        }}
}



\author{
    \IEEEauthorblockN{
        Robert~A.~Bridges\IEEEauthorrefmark{2}, 
        Brian~Weber\IEEEauthorrefmark{2},
        Justin~M.~Beaver\IEEEauthorrefmark{3},
        Jared~M.~Smith\IEEEauthorrefmark{5},
        Miki~E.~Verma\IEEEauthorrefmark{1},\\
        Savannah~Norem\IEEEauthorrefmark{2}, 
        Kevin~Spakes\IEEEauthorrefmark{2}, 
        Cory~Watson\IEEEauthorrefmark{2}, 
        Jeff~A.~Nichols\IEEEauthorrefmark{2},
        Brian~Jewell\IEEEauthorrefmark{2},\\
        Michael.~D.~Iannacone\IEEEauthorrefmark{2},
        Chelsey~Dunivan~Stahl\IEEEauthorrefmark{2},
        Kelly~M.T.~Huffer\IEEEauthorrefmark{2}, 
        T.~Sean~Oesch\IEEEauthorrefmark{2} 
        }
    \IEEEauthorblockA{
        \IEEEauthorrefmark{2}Oak Ridge National Laboratory, Oak Ridge, TN
        \IEEEauthorrefmark{3}Lirio LLC, Knoxville, TN\\
        \IEEEauthorrefmark{5}SecurityScorecard, New York, NY
        \IEEEauthorrefmark{1}Stanford University, Palo Alto, CA\\
        \\
        \{bridgesra, weberb, spakeskd, watsoncl1, jewellbc, iannaconemd, nicholsja2, dunivanck, hufferkm,  oeschts\} @ornl.gov} 
        jbeaver@lirio.com, jared@jaredsmith.io, miki@rewiringamerica.org, savannah.norem@gmail.com
    }



\maketitle

\thispagestyle{plain}
\pagestyle{plain}

 \begin{abstract}
 This work presents an evaluation of six prominent commercial endpoint malware detectors, a network malware detector, and a file-conviction algorithm from a cyber technology vendor.
  The evaluation was administered as the first of the Artificial Intelligence Applications to Autonomous Cybersecurity (AI ATAC) prize challenges, funded by / completed in service of the US Navy.
The experiment employed 100K files (50/50\% benign/malicious) with a stratified distribution of file types, including $\smalltilde$1K zero-day program executables (increasing experiment size two orders of magnitude over previous work).
We present an evaluation process of delivering a file to a fresh virtual machine donning the detection technology, waiting 90s to allow static detection, then executing the file and waiting another period for dynamic detection; this allows greater fidelity in the observational data than previous experiments, in particular, resource and time-to-detection statistics. 
To execute all 800K trials (100K files $\times$ 8 tools), a software framework is designed to choreographed the experiment into a completely automated, time-synced, and reproducible workflow with substantial parallelization.
Software with base classes for this framework are provided. 
A cost-benefit model was configured to integrate the tools' recall, precision, time to detection, and resource requirements into a single comparable quantity by simulating costs of use. 
This provides a ranking methodology for cyber competitions and a lens through which to reason about the varied statistical viewpoints of the results. 
These statistical and cost-model results provide insights on state of commercial malware detection. 
\end{abstract}


\section{Introduction} 
\label{sec:intro}
Malicious software or malware refers to any file that seeks to disrupt operations, corrupt data, allow unauthorized access to data or systems, or otherwise cause unwanted consequences to users. 
Given the repeated reports of widespread use of malware, 
antivirus (AV) software or endpoint malware detection technologies---which reside on each host and provide detection and quarantine capabilities against malware---are one of the most common and critical security defenses. 

Academic research on the efficacy of commercial malware detectors is fairly nascent---see the Related Works \ref{sec:related-works}. 
Notably, evaluations provided by industry exist but are either small in terms of test sample size (e.g., \cite{mitre}) or lack enough difficulty in the testing samples to differentiate tools (e.g., \cite{av-test, selabs-test}). 
As a result, organizations have limited insight into whether or how commercial malware detection tools add value to their current defense solutions, short of trusting possibly skewed vendor-provided claims and market/consumer reports. 
To illustrate, we highlight recent results. 
In April 2022, AV-Test \cite{av-test}, a commercial company providing large scale tests on malware detection technologies, presents results in which all twenty evaluated malware detection technologies received a score at least 5.5 of 6 possible  (17 were rated perfectly 6/6), and claim 99.8\% recall on zero-day files as an industry average. 
Tests from SE Labs {\cite{selabs-test}} in early 2022 report approximately ten malware detectors that exhibit above 99\% accuracy in most tests. 
Such results may, at first glance, lead one to believe that malware detection is ``a solved problem'' thanks to our commercial technologies.
We contrast these promising results with recent academic literature. 
The most recent research evaluation of commercial malware tools is Bridges et al. \cite{bridges2022beyond}, which tested four commercial malware detectors with $\smalltilde 3.5$K samples. 
Their experiment revealed: 
    near perfect precision in all four tested tools; 
    recall of \smalltilde 35\%--55\%,  with higher recall for program executable (PE) file type, dropping for many other file types;
    that machine learning (ML)--based tools achieve mutually similar detection rates (\smalltilde 40\%) on zero-day (never-before-seen) files while signature-based tools fail to detect nearly all zero-days (\smalltilde 4\% detection rate).
Another recent study of Zhu et al. \cite{zhu2020measuring} leveraged VirusTotal (\url{www.virustotal.com}, an online threat intelligence website that provides many commercial detection engine's results on a submitted file) to evaluate detectors on 120 zero-day ransomware samples and 236 benignware samples finding true \textit{and false positive} results varying from  25--100\%. 
Notably, Zhu et al. also found that VirusTotal  versions of commercial detectors have on average higher recall but more false positives than their endpoint products. 
Finally, we invite readers to peek ahead at the summarized findings of this research effort---itemized at the end of this introduction, Section \ref{sec:edr-takeaways}---which support the narrative that commercial detectors have (perhaps alarmingly) room for improvement. 

Overall, academic research paints a \textit{much} less rosy picture of current malware defenses than the commercial evaluations. 
As commercial malware detectors are a critical line of defense, development of rigorous evaluation methodologies and dissemination of malware detection experimental results are needed to both understand our vulnerabilities and to help prioritize future research.

Seeing this need, the US Navy commissioned a series of three challenges (see AI ATAC 1 \cite{ai-atac1-website}, AI ATAC 2 \cite{ai-atac2-website}, AI ATAC 3 \cite{ai-atac3-website, bridges2023testing})
This paper documents the experiment and results that constitute the AI ATAC 1 Endpoint Malware Detection Challenge. 
Using 100K files to test eight malware detectors---six endpoint detectors, one network detector, and a malware detection algorithm provided by a security vendor,
our experiment is the largest in terms of numbers of files, research evaluation  of commercially available malware detectors,. 
As an agreement of the challenge, we must hold anonymous the technologies evaluated. 
All were from prominent vendors in the cyber technology market.

When analyzing efficacy of malware detectors, there are more nuanced questions than overall detection statistics, e.g., accuracy, recall, false positive rates, that are important for both research prioritization and for the selection/use of modern detectors.
Machine-learning promises greater defenses to zero-day threats, yet, in our context of commercial malware detection, comparisons and experimentation to understand the fulfillment of this promise is needed; furthermore commercial malware seems focused mainly on program executable (PE) file type, as evidenced by previous works \cite{apruzzese2022role, bridges2022beyond, zhu2020measuring, koch2022toward}. 
We provide research results comparing signature- and machine-learning-based tools on a set of zero-days and provide results conditioned on filetype using a corpus with varied filetype distribution.


\begin{figure}[ht]
    \centering
    \includegraphics[width=0.49\textwidth]{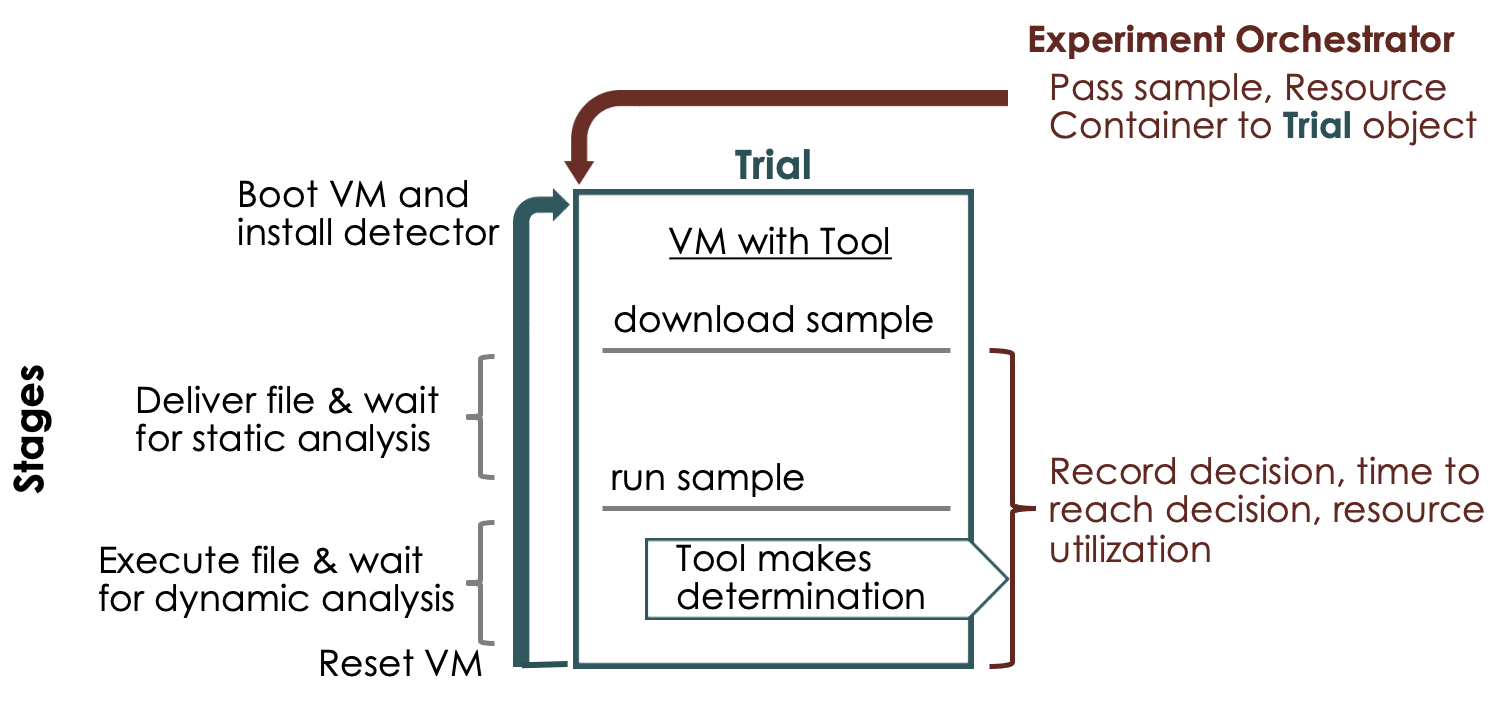}
    \caption{Trial workflow depicted.
    In practice, the Experiment Orchestrator ran \smalltilde 2K Trials in parallel.}
    \label{fig:trial-workflow}
\end{figure}

Descriptions of the design and configuration of the research range (data center) built to accommodate this and other experiments is detailed in our previous work \cite{nichols2021assembling}.
Our experiment is influenced by the previous work of Bridges et al. \cite{bridges2022beyond}, but contains the following novel contributions.  
Our work doubles the number of prominent commercial detectors tested; 
expands the number of files by two orders of magnitude; 
increases fidelity of the experimental observations by including host resource observations; 
and adds fidelity to a previously used cost-benefit analysis method.  
This paper presents a novel experimental workflow for gathering resource and file conviction data when testing malware detectors on both unexecuted files and during/after execution, an advancement in the testing methodology used by previous works. 
Lastly, this paper provides significant engineering advancements---a general software framework to repeatably automate the experimental workflow with substantial parallelization---over the previous research literature to accommodate the scale (100K files $\times$ 8 detectors). 
These contributions are discussed in greater detail in the following subsections.

\subsubsection*{Experimental Design}
The design of the experiment provides a higher fidelity evaluation for many reasons, besides the increase in file quantity. 
We define a \textit{trial} as a sub-experiment testing  a single tool's capability to test a single file. 
Each trial progressed through \textit{stages}: 
first the file was delivered to a fresh VM instantiated with the detection technology; then after \smalltilde 90~s the file was executed; finally after a period of time the VM was closed 
(see Figure \ref{fig:trial-workflow}.)
These stages allowed for the tool under test to detect the file if not statically (before execution) then dynamically (during or after execution). 
At the cost of much greater computational expense, this provides higher fidelity in the experiment with respect to how a detector will react when presented with a file on a host (e.g., our Baseline 2 tool seemingly attempts to analyze a file only once it is executed) and therefore allows more accurate time to detection, resource use, and accuracy observations.
Resource utilization of the tool under test was recorded and taken into account in this study (which was not done in previous studies). 
In our corpus we took care to try to match the percentage of files of each type to the distribution reported by VirusTotal to gain realism. 
With the help of a security operation center (SOC), we used real statistics on the number of files encountered per year by an endpoint detector to inform our cost-benefit analysis.  

\subsubsection*{Experiment Implementation Framework}
As a second contribution, we discuss a novel software framework for repeatably implementing the experimental workflow, allowing scalability to accommodate the dynamic experimental workflow by facilitating thousands of parallelized trials. 
Given that the experiment required 800K trials, 
a simple serial implementation, taking an estimated \smalltilde 4.5 years,  was completed in \smalltilde 15 hours per tool or \smalltilde 5 days in total.

\subsubsection*{Cost-Benefit Analysis}
Although the scale 
allowed for more fine-grained results (recall, precision, time to detection, resource metrics), a scoring framework that translates the many experimental observations to a real-world context was needed to reason about the tools' capabilities and, for the sake of the competition, to make the multidimensional measurements comparable. 
A general cost-benefit analysis \cite{iannacone2020quantifiable}  was adapted to this experiment. 
There are two  benefits of using the cost-benefit analysis framework:
(1) the model integrates the many experimental measurements into a single, comparable, quantifiable cost by simulating how the tool would perform in a security operation and  
(2) it provides a framework for quantifiably reasoning about the many different statistical results.
Our implementation of the cost model follows previous works
with two additions: 
it is augmented to accommodate the higher-fidelity experimental results used in this paper;
it employs data observations from a SOC to understand the quantity of file decisions a host makes in its first year. We illuminate drawbacks to the cost model that contribute to future research.


\subsubsection*{Commercial Malware Detectors Summarized Findings}
\label{sec:edr-takeaways}
Our final contribution are the results of the commercial  detectors from this experiment 
All tools provide near perfect precision and, except for the lone dynamic, network-based malware detector,  achieve recall of \smalltilde 50\% (on corpora of 98\% public malware files). 
 On public malware, signature-based detectors are competitive if not better in recall and time to detection than host, ML-based detectors.
On never-before-seen files, signature-based detectors fail almost categorically (\smalltilde 4\% recall  on zero-day malware), while ML-based tools provide about 10$\times$ in recall.
Our cost model analysis confirms the previous two bullets with a signature-based tool prevailing when costs per malware are constant, but ML-based tools will prevail if one assumes zero-day files accrue much greater costs. 
The lone dynamic detector has the highest recall.
Many tools detect only PE and possibly MS Office file types, while other tools provide competitive recall across all tested file types. There is large variance in capabilities on non-PE file types.  
Time from file delivery until detection is only a matter of seconds (in median) for most tools, with one tool taking median \smalltilde 1m, and one tool waiting until file execution to begin analysis, then taking 9s.  

\subsection{Related Works}
\label{sec:related-works}
We focus on evaluations of commercial malware detectors, as these are the directly related works. 
Adjacent lines of research include economic simulations for cyber operations
and cyber competition design and scoring
both surveyed by Iannacone \& Bridges \cite{iannacone2020quantifiable}.

To investigate what an adversary can learn about a blackbox malware detector, Christodorescu \& Jha (c. 2004) \cite{christodorescu2004testing} evaluate three commercial malware detectors using only eight malicous samples, finding dismay at the ``dismal'' state of malware detectors. 
Pandey \& Mehtre \cite{pandey2014performance} (c. 2014) compute accuracy, recall, specificity, and a poorly described ``efficiency'' quantity for  17 different file analysis or online services using 29 malicious samples. 
Aslan et al. (c. 2018) \cite{aslan2017investigation} use 100 benign plus 100 malicious to evaluate many static, dynamic, and online file analysis tools. 
Detection rates vary from 62\%--77\%, with dynamic analysis tools outperforming static and combinations of tools providing better coverage than any single member of the ensemble. 
Fleshman et al. (c. 2018) \cite{fleshman2018static} perturb known malware to create 1K total obfuscated  samples. Commercial malware detectors and noncommercial supervised algorithms are tested. 
The two ML algorithms are found to be more robust to perturbations than commercial detectors. 
Zhu et al. (c. 2020) \cite{zhu2020measuring} investigate the benign/malicious convictions of \smalltilde 70 commercial malware detection engines on over 14K files daily for more than a year via VirusTotal.   
Zhu et al. show that using thresholded voting  (  $\geq n$  vendors convict $\implies$ malicious) is quite accurate and robust to each detector's day-to-day decisions, which are volatile.
Further analysis using 60 obfuscations of two known ransomware (120 zero-day malware) and 256 benign samples shows wide variation in recall and precision, and that many commercial detectors exhibit results much lower than our study finds (See Zhu et al.'s fig. 13). 
Notably, Zhu et al. continue to compare results of 36  endpoint counterparts with the VirusTotal detection engines and find that that they exhibit greater recall but worse precision than their online counterparts. 

By far the closest work to, and indeed the building block for this work, is Bridges et al. \cite{bridges2022beyond} (c. 2022), which investigated two popular commercial host-based malware detectors---one signature-based and one ML based---and two popular commercial network-level detectors---both ML based, one using static analysis, and one using dynamic analysis---with a test corpus of \smalltilde 3.5K files. 
Bridges et al. focus on comparing ML-based detectors to signature-based detectors using zero-days (as does this work) and are the first to tailor the general cost-benefit framework introduced by Iannacone \& Bridges \cite{iannacone2020quantifiable} for a file conviction experiment. 
Results showed detection rates in the 34\%--55\% range, with any pair of network and host detectors (logical \texttt{OR} of both tool's detection results) achieving \smalltilde 60\% recall. 
The signature-based detector is superior if all malware (i.e., zero-days and public malware) accrue  cost identically; yet, ML-based tools win if the maximum zero-day attack cost (an input parameter) far outweighs the maximum  public malware costs. 
Bridges et al. go on to reconfigure the cost model to simulate savings of adding a network detector to a given host-based detector, concluding that (under the experiment's assumptions) adding a network malware detector will save money after the first year. 
Notable differences from our current work are that Bridges et al. used only \smalltilde 3.5K files, with imbalanced classes (78\% malware), delivered over multiple protocols in a test network to enable the network-level detectors, and did not execute files to enable dynamic detection. 
Our current work scales the experiment to 100K files, provides balanced classes (50/50\%),
matches the file type distribution in our test corpus to the ``real-world'' sample observed by VirusTotal as closely as possible, incorporates host resource usage by the detector as observations, executes files on the host, and advances the cost model to incorporate the added observations. 

\section{AI ATAC 1 Malware Detection Competition}

AI ATAC 1 focused on evaluating endpoint malware detectors and required use of ML for the file conviction \cite{ai-atac1-website}.
Notably, signatures can be used alongside or in a pipeline with ML. 
By ``endpoint malware detector'' we mean the technology should reside on a host, automatically identify the presence of a file on the host, and provide malware classification for the file.

\subsubsection*{Tools and Samples Used} 
Four commercial vendors submitted ML-based, endpoint malware detection engines to the competition, and two signature-based tools were used as baselines for comparison. 
For comparison, we also test two other detectors: (1) a popular commercial network-level malware detector that claims to use dynamic analysis in virtual environments to  build feature vectors before applying supervised learning and  
(2) a pretrained supervised learning malware detection algorithm provided by a popular cyber tech company in the form of a software development kit, using the same 100K file test corpus. 
Because these two tools are not host endpoint detection tools, the cost model, as configured for endpoints cannot be directly compared. Similarly, timing results per file are not comparable. On the other hand, the statistical detection results are comparable and are provided to gain awareness of the state of the art in practice and provide context for the endpoint detectors' capabilities. 
All tools were installed and configured with vendor support. ML-based tools were pretrained before submission. 

\begin{figure}[ht]
    \includegraphics[width=.49\textwidth]{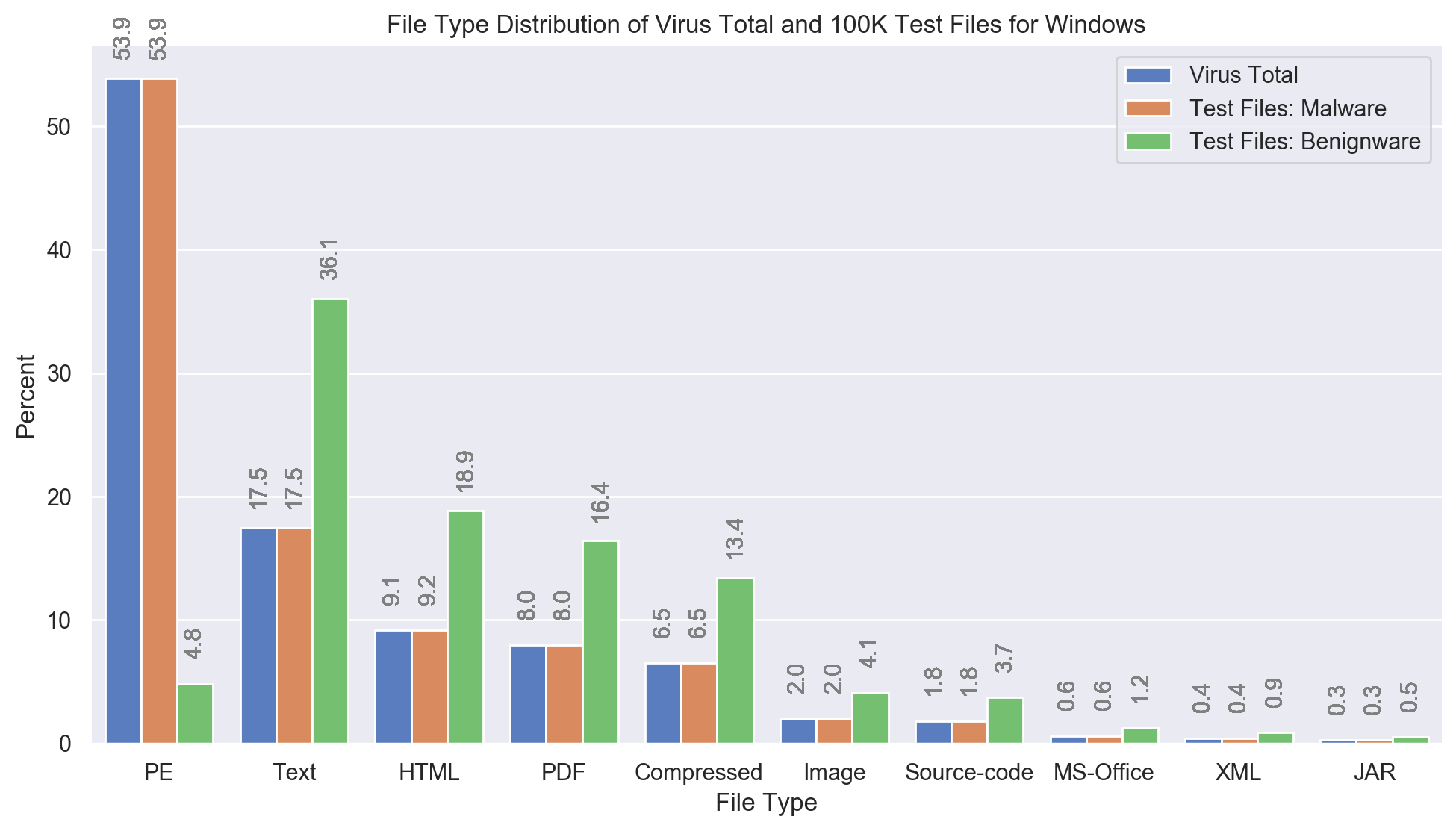}
    \caption{Distribution of file types in the test corpus versus VirusTotal. A corpus of 100K test files (50/50\% malware/benignware) was compiled for the AI ATAC 1 Endpoint Malware Detection Competition. This bar chart shows the VirusTotal distribution of file types, considered in this work to be the ``real-world'' distribution of file types, against the distribution of  malware and benignware used in the competition.}
    \label{fig:file_dist_windows}
\end{figure}

To collect benignware, we leveraged websites such as Softpedia (\url{https://www.softpedia.com}), FileHippo (\url{https://filehippo.com/}), and the Maven (\url{https://maven.apache.org/} \cite{maven-jars-tutorial}) repository of Java archives. 
We automated the scraping of these websites to collect what we deemed benignware. 
We used these sites because their popularity, along with the popularity of the downloads they host, lends them to trustworthiness. For example, kommandotech (\url{https://kommandotech.com/statistics/chrome-usage-statistics/}) puts global Google Chrome users at 2.94 billion, with the software being updated roughly every six months. 
After eliminating duplicates, we had 1,049,981 benign samples, consisting mostly of PE, text, HTML, PDF, and compressed files. 
We then took a subsample of these files to reduce our selection to 50K benign files that matched the VirusTotal distribution as closely as possible---discrepancies between the VirtusTotal distribution of file types and  our benign file type distribution exist because we were unable to obtain enough benign files of certain types (Figure \ref{fig:file_dist_windows}). 
Most malicious files were obtained through VirusShare (\url{www.virusshare.com} and a collection of about 1K zero-days were provided by and described in Bridges et al. \cite{bridges2022beyond};
After gathering our samples, we used VirusTotal to get data on how often specific file types show up as malware. 

\subsection{Experimental Workflow} 
The experiment was designed to test the detectors' capabilities when instantiated on an endpoint that receives a new file. 
To implement a \textit{trial}---testing a particular detector in the presence of a particular file---a VM was instantiated with a fresh Windows OS and the detection technology and was configured to forward any alert to an out-of-band experiment orchestration node.
The experiment orchestration node would progress through the following \textit{stages}: 
(1) deliver the test file to the VM and wait an average of 90 s, then 
(2) execute the file, wait another 90 s, and 
(3) close the VM.  
Timestamps for these three events were recorded and could be compared with the timestamps of the alert logs to identify the time to detection and whether the file was executed before detection occurred. 
Further, the experiment orchestration node collected statistics on the resource needs of the malware detector under test;  specifically, CPU (percentage), RAM (bytes), and disk I/O (bytes read, bytes written) were programmatically collected. 
Network bandwidth was desired, but our experimental design was unable to isolate only the network resources used by the detector.  
Independent of any file, for each tool, the same resource measurements (CPU, RAM, disk I/O) were recorded for five  minutes of the detection process sitting idle. 
This was needed as a baseline.

\subsection{Parallelized Experimental Framework} 
Altogether, this experiment involved running this test workflow for 800K trials (100K files for all eight tools) with an average of 180s per trial. 
Implementing a repeatable experiment of this scale required orchestration software engineered to run many simultaneous, identical tests. 
To prevent malware infections from affecting results, per-file instantiations of VMs, detection tools (under test), and the resource and accuracy monitoring capabilities were needed. 
To execute all 800K trials in a serial implementation, about 4.5 years would be needed; hence, a large focus of our research was on designing and implementing a software framework that choreographed the whole experiment into a completely automated and reproducible workflow with substantial parallelization. 
Here we describe this software engineering feat. 
With our hardware (see previous work of \textcolor{ForestGreen}{Redacted Citation} 
for details), we came up with a design where a single orchestration node would asynchronously conduct \smalltilde2,000 virtual machines, guiding them in parallel through the steps to test a single sample and moving them each on to the next stage of the trial as they finish. This reduced the experimental time to 15 hours per tool and 
importantly allowed experimental results that fit the timeline of the competition.

\subsubsection{Design (with Dos and Don'ts)} 
Our initial effort for a parallelized framework used one process to manage one VM (trial). 
Timing experiments were run to gain data on the performance of simply creating $n$ consecutive process, for $n = 1, ..., 1000$. Notably, the growth was nonlinear, seemingly quadratic, and simple regression, which yielded an extrapolated estimate of 12 hours to simply create 10K processes. 
Further experiments on the overhead latency encountered when using a one-to-one relationship between processes and trials steered us to a many-trials-to-one-process design. 
Attempting the opposite extreme, one orchestration process for all trials, revealed its own unique hurdles. 
We found that using one process to orchestrate all 2K simultaneous trials took on the order of tens of seconds to check the status of every trial due to compounding network latency. This led to a non-negligible amount of VMs sitting idle or being powered off at any given point in time while waiting for the orchestrator to start the next stage of the trial. 
Ultimately, we settled on a balanced approach in which 100 orchestrator processes handled \smalltilde 20 trials each, mitigating the drawbacks of either extreme.

\subsubsection{Framework Objects}

\begin{figure}
    \centering
    \includegraphics[width=.485\textwidth]{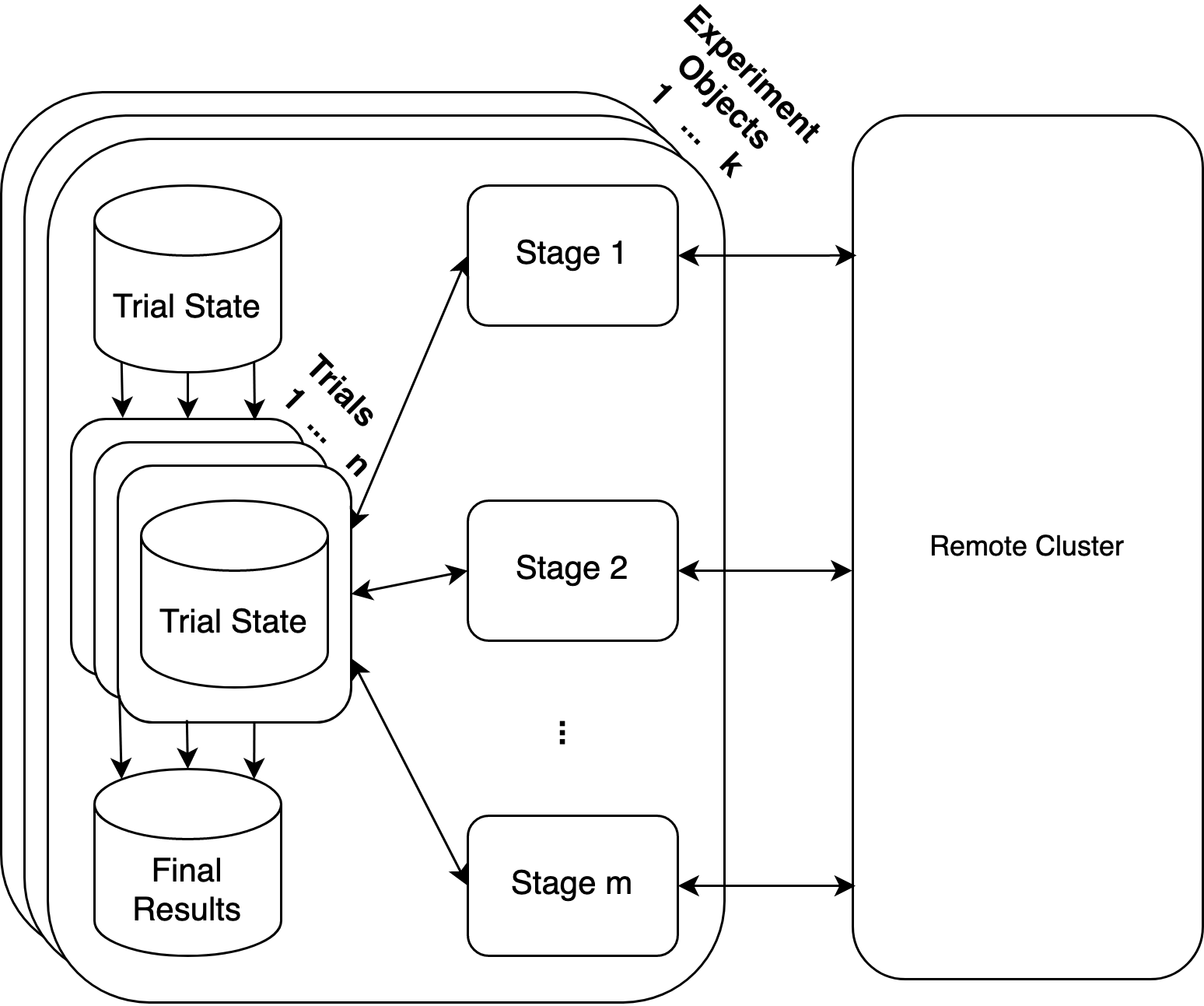}
    \caption{Experiment objects run in parallel on a local machine, and each runs many Trials in parallel. Experiment objects iterate over Trials, pass them Trial parameters, and start their first Stage. After each Trial’s first Stage has started, the Experiment object continues iterating over Trials, giving each a chance to check on the status of its current Stage and advance Stages if the current Stage is finished. Stages contain methods for starting and checking the status of Trial Stages on a remote resource based on Trial state. The Trial state might include Trial parameters, temporary data, and intermediate results. When a Trial has finished all of its stages, the Experiment object stores the final results from the Trial and passes the Trial the next set of parameters to test, thereby starting the next trial. This continues until all parameters are exhausted. Not pictured are the resource management classes (ResourceWarden and ResourceContainer) and generic Exception handling, which are held as attributes of the base classes.}
    \label{fig:experiment_framework}
\end{figure}

Our framework, see Figure \ref{fig:experiment_framework}, uses several \textbf{Experiment Orchestrator} objects, which run in parallel and are given a list of trials to execute with their parameters and, together, run the whole experiment.
The Experiment Orchestrators: hare a \textbf{Resource Warden} object that is passed the list of resources and holds a dictionary of semaphores (each representing a resource) used for load balancing; 
instantiate a \textbf{Resource Container} object that requests resources from the shared Resource Warden, tracks its held resources, and releases resources back to the warden; 
instantiate, in parallel, many \textbf{Trial} objects and pass the Trial objects their respective parameters along with the Resource Container, which allows the Trials to acquire or release resources if necessary; 
receive and handle \textbf{Trial Results} object upon completion of each trial;
reset the finished Trial objects and pass them the next set of trial parameters until the list of trial parameters is exhausted and results are obtained; and
collect performance metrics and logs them to Kafka to monitor experiment runs. 
A separate process consumes these logs and displays aggregate performance statistics, such as the number of files tested per minute, number of failed runs (if any), etc.

\textbf{Trial} objects are responsible for running every stage of a trial and returning results. 
The Trial objects: contain a list of references to each \textbf{Stage} object (in order); 
run, in serial, each Stage object;
check and store the current stage status, data, and results; 
pass a Trial Results object to the Experiment Orchestrator upon finishing the trial.  


\textbf{Stage} objects contain methods for starting and checking on the status of individual steps in a Trial and are not intended to store any data.
The first stage to testing an individual file is to restore a VM to a live snapshot. 
The next step is to upload a script to the VM and execute it. 
The script that is uploaded is executed and passes a URL to an internal webserver to download the sample of malware to be tested. The script then waits one minute to give the tool time to statically detect the malware. 
After one minute, the sample is executed to give the tool to dynamically detect the sample. 
The script then waits another minute before finishing. 
Note that latency in execution resulted in \smalltilde 90s, a 30s delay. 
In addition, the script collects all resource usage from each of the malware detection tool's processes and logs it to be downloaded later. 
In addition, exit codes, standard output, and standard error from the executed malware are logged. 
Finally, the last Stage downloads the results files from the VM containing all the results and powers it off to save host resources until it is needed again. 
If possible, detection results from the tool under test were pulled from logs local to the VM and reported in the results immediately. 
Otherwise, detection results were downloaded from the tool's management interface and merged with other results after the full experiment was run.

After scaling, this setup created some issues because of the workload being generated on the VMWare cluster. 
We discovered VMWare is not designed for this use case and does very minimal load balancing on its own (e.g., restoring a VM from snapshot consumes a large amount of disk I/O). 
Consequently, trying to restore more than 2,000 VMs to a snapshot at one time generated induced stability issues and very significant degradation in performance. 
Stability, along with the need to provide sufficient resources to tools, are the motivations for load balancing---the Resource Warden and Resource Container objects. 

Load balancing greatly improved throughput and stability by limiting the amount of concurrent disk-heavy VM operations, but a small amount (\smalltilde10 per tool) of unexpected VMWare errors still occurred.
To handle these remaining errors, we created function decorators and custom \textbf{Exception} classes to catch exceptions and apply selected generic remediations.
Generic responses to exceptions include: 
ignoring the exception and continuing, 
restarting the current stage, 
 skipping the current stage, 
 restarting the current trial,
 aborting the current trial without collecting results, 
 restarting the Experiment Orchestrator object, and
 aborting the experiment. 
To select responses, the Experiment Orchestrator is configured by providing a list of responses to apply to exceptions, in order, at the Stage, Trial, and Exception level. 
For example, Stages can be configured to retry the current stage twice at the stage level. 
If that does not resolve the issue, the Trial can be configured to restart at the Trial level. 
Stage and Trial exception responses reset after the successful completion of that Stage or Trial. 
Our configuration ignores exceptions three times before trying to reset the stage between zero and two times depending on the stage. 
If none of the stage responses work, Trials are configured to reset three times before aborting the Trial and throwing out the results.
After adding load balancing and generic error handling, we were able to complete the testing of 100K files in $\smalltilde$15 hours without loss of results due to VMWare API Exceptions, all while providing sufficient resources to VMs running the tools.

\subsubsection{Technical Drawbacks}
\label{sec:technical-drawbacks}
There are several issues to our approach that we were not able to address due to time and resource constraints. First, we were unable to rigorously tune the ResourceWarden configuration. The ResourceWarden was configured via trial and error to limit the number of VMs performing certain steps at the same time. E.g., only four VMs across the cluster were allowed to be restoring from a snapshot at any given time due to hardware limitations. More formal tuning may have allowed higher throughput and/or lower resource usage.

Next, it is well known that many malware samples hide their malicious behavior to avoid analysis if it is not executed in a correct environment or using a correct method. Due to the significant technical challenge it would take to address this issue for all 50K malware samples, minimal effort was given to make sure malware executed as intended. Many commercial and open source sandboxes go through significant effort to address this issue, however development time to replicate and integrate these solutions was highly prohibitive in our case. To conduct dynamic analysis, we simply opened the malware using a reader for the associated file type, which allowed the tool to observe its behavior. For example, we executed Windows PE files with no arguments, opened PDF files in Adobe Reader, and opened office documents in Microsoft Office.

While we originally intended for each sample to sit statically on the VM for 60s, the average was $\smalltilde$ 90s. Since each orchestration process checks the status of its assigned VM in a loop, if a VM finishes the static detection right after the orchestration node finishes checking it, it has to wait until every other VM is checked again before it moved on to the next stage, dynamic detection. We ultimately included detections during this time as a positive label, since many tools automatically removed or quarantined detected files. Including them as a negative label would have unfairly penalized these tools.

    \begin{table*}[ht]
\centering
\begin{threeparttable}
\caption{Average Resource Costs (Top) and Average Detection Costs (Bottom) Per Malware and Benignware}
\label{tab:ave-costs-per-file}
\begin{tabular}{lllllll}
\toprule
                             & \textbf{Tool 1}               & \textbf{Tool 2}             & \textbf{Tool 3}             & \textbf{Tool 4 }            & \textbf{Baseline 1}           & \textbf{Baseline 2}            \\
\cmidrule{2-7}
Ave Benignware Resource Cost & \$ 0.002596  & \$ 0.002594 & \$ 0.013464 & \$ 0.013959 & \$ \textbf{0.002279} & \$ 0.003078  \\ 
 {Ave Malware Resource Cost}    & \$ 0.002987 & \$ 0.002632 & \$ 0.012398 &  \$ 0.015949  & \$ \textbf{0.002543} & \$ 0.003276 \\
\cmidrule{2-7}
 {Ave Benignware Detect Cost} & \$ 0.136695   & \$  0.032947   & \$ 0.023133 & \$ 0.012618   & \$ \textbf{0.003505}  & \$ 0.007711   \\
 {Ave Malware Detect Cost}    & \$ 559.932        & \$ 533.515    & \$ 494.544   & \$ 566.961       & \$ 614.967   & \$ \textbf{353.252} \\
\bottomrule
\end{tabular}
\end{threeparttable}
\end{table*}

\subsection{Scoring Framework for AI ATAC 1} 
In all we have four documents of raw data to be used for the cost model scoring:
\begin{myitemize}
  \item File information: A table of 100K files used for testing with filename, file type, and label (malicious/benign); 
  \item Measurements from the tests: A table providing for each tool and file; the experiment produced  timestamps for file download and execution, the resource observations (CPU, RAM, HDD) for that trial.
  \item Alert outputs: A table curated from each tool's alert logs providing the file name, alert, time-synced timestamp. 
  \item Ambient resource data: A table with ambient CPU (percentage), RAM (bytes), and HDD (bytes) per tool. 
\end{myitemize}
  
Standard statistical results for precision, recall, F1, and time-to-detection are presented; in particular, comparisons of detection statistics between zero-day to public PEs are examined.  
For the AI ATAC competition, we require a single, comparable quantity to rank the detectors; hence, for each tool, we will simulate a cost incurred for the first year on one host. This provides a single lens through which to integrate and reason about about the many measurements. 

\textbf{Initial Costs---}Each tool's subscription fees, setup, and/or hardware needed was estimated.
Based on our setup experience, if a tool requires  more than eight labor hours, we added that time into the initial costs at a rate of \$70/hour, the SOC operators fully burdened cost used in previous work \cite{bridges2022beyond}. 
Tools that require an on-premises appliance, database, etc., have a cost added to account for the added hardware, subscription, electricity, and labor during setup. 
We assume a 10K IP network and divide the estimated total appliance costs by 10K as we are estimating costs for 1 host.

\textbf{Ongoing Resource Costs---}We leverage rates provided by cloud service providers for CPU, RAM, and HDD costs. 
In particular, we based our rates on a recent publication of Dreher et al. \cite{dreher2016cost}. The resources on the host are charged cloud costs divided by three (to
    remove profits made by cloud costs), namely: 
\$0.02444/3 per hour for full CPU, 
\$.00328/3 per hour for 1 GB RAM,
\$0.05/3 for 1 GB HDD/month $\times$  1 month/30.5 days $\times$ 1 day/24 hours. 
Because we could not measure labor for tuning tool and reconfiguring
 tools over time, this ongoing real-world cost is not taken into account in this model. From the 5m of ambient resource observations, we compute the average
  ambient resource cost for a minute.

For each tool we compute the average resource costs per benign file and per malware, which are reported in the top of Table \ref{tab:ave-costs-per-file} and will extend these linearly to an estimated cost for the year. 
To this end, we accrue the resource costs for every tool and for every file used in the experiment according to the costs in the preceding paragraph. 
We cannot use the resource measurements from the time of file execution until the VM is closed because these measurements include both the detector's and the file's resource needs. 
Hence, we use the resource costs from the delivery of the file until the execution, then we linearly scale the ambient resource costs for that detector to account for the resources used from the time of execution until detection.
If there is a file for which the resource observations were not present in the data (small minority), we use the average ambient resource costs to impute the resource cost.

\begin{figure} 
\centering
\includegraphics[width = .53\textwidth]{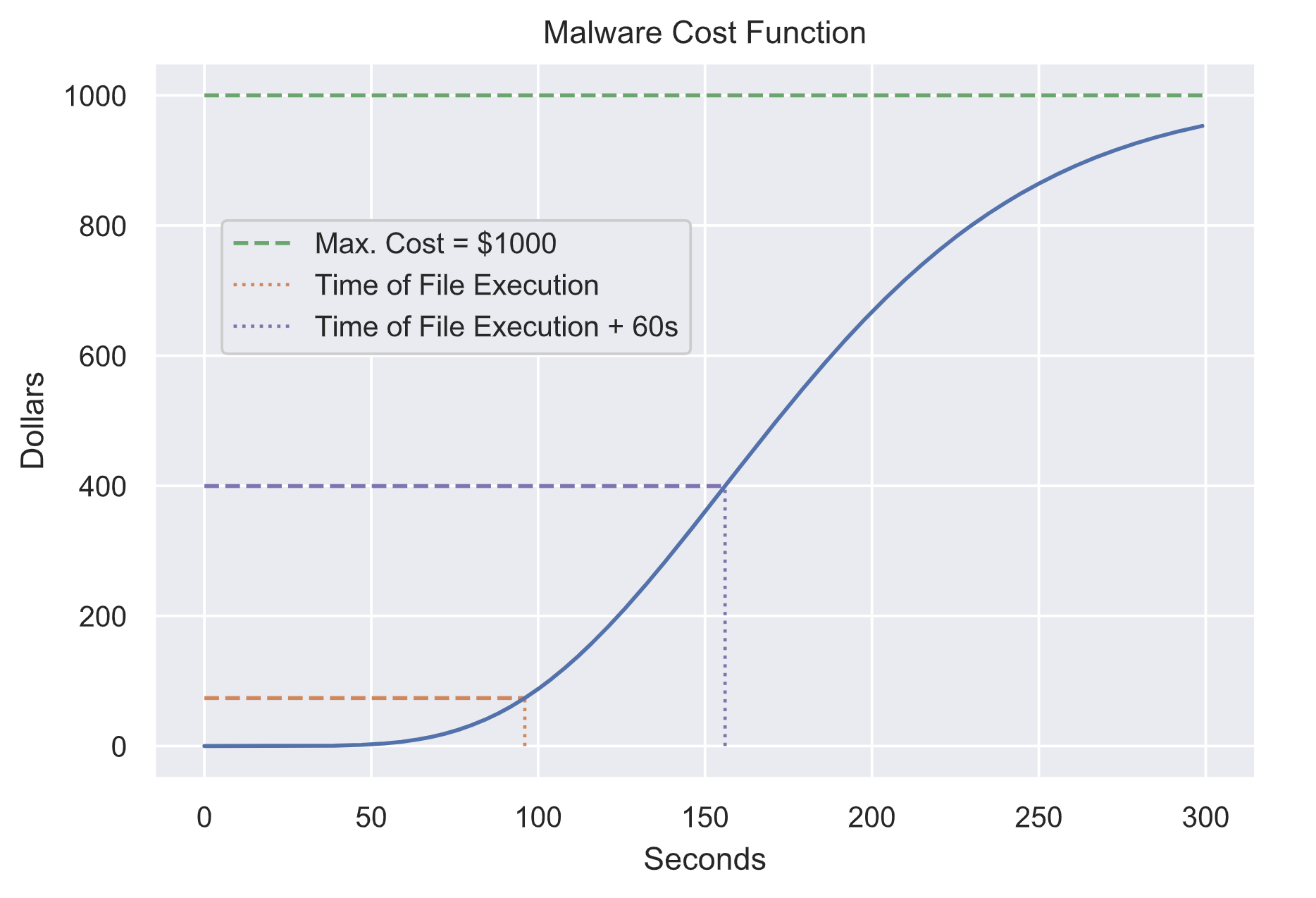}
\caption{Attack cost S curve} 
\label{fig:attack-cost}
\end{figure}

\begin{table*}
\centering
\caption{Cost Model Results with Subtotals (Top) and Detection Statistics (Bottom)}
\label{tab:results}
\begin{threeparttable}
\begin{tabular}{lllllllll}
\toprule
    & \textbf{Tool 1} & \textbf{Tool 2}  & \textbf{Tool 3} &  \textbf{Tool 4 } & \textbf{Baseline 1} & \textbf{Baseline 2} & \textbf{Network} & \textbf{Algorithm}           \\
\cmidrule{2-9}
\rowcolor[HTML]{EFEFEF} 
\textit{Initial Cost }           & \$ 39   & \$ 10  & \$ 12  & \$ 8  & \$ -~~ & \$ -~~ & n/a & n/a \\
  \\
\rowcolor[HTML]{EFEFEF}
Annual Malware Resource Cost          & \$ 2   & \$ 2 & \$ 7  & \$ 9  & \$ 1 & \$ 2   & n/a & n/a \\
Annual Benignware Resource Cost       & \$ 128       & \$ 128   & \$ 665   & \$ 690    & \$ 113 & \$ 152 & n/a & n/a \\
\rowcolor[HTML]{EFEFEF}
Annual Ambient Endpoint Resource Cost & \$ 814       & \$ 439   & \$ 2,618   & \$ 2,337    & \$ 453  & \$ 842    & n/a & n/a \\
Annual Appliance Resource Cost        & \$ -  & \$ 10  & \$ 2 & \$ - & \$ - & \$ -  & n/a & n/a \\
\rowcolor[HTML]{EFEFEF}
\textit{Annual Resource Cost     }            & \$ 944       & \$ 569   & \$ 3,291     & \$ 3,036    & \$ 567 & \$ 996    & n/a & n/a \\
\\
\rowcolor[HTML]{EFEFEF}
Annual Malware Detect Cost            & \$ 324,760             & \$ 309,439      & \$ 286,835     & \$ 328,837     & \$ 356,681  & \$ 204,886       & n/a & n/a \\
Annual Benignware Detect Cost         & \$ 6,755       & \$ 1,628    & \$ 1,143     & \$ 624  & \$ 173 &  \$ 381    & n/a & n/a \\
\rowcolor[HTML]{EFEFEF}
\textit{Annual Detect Cost  }  & \$ 331,516             & \$ 311,067      & \$ 287,978           & \$ 329,461   & \$ 356,854    &  \$ 205,267       & n/a & n/a \\
\\
\rowcolor[HTML]{EFEFEF}
\textbf{Total Cost}      & \$ 332,499            & \$ 311,636      & \$ 291,279  & \$ 332,505       & \$ 357,422 & \$ \textbf{206,263}       & n/a & n/a \\
\cmidrule{2-9}
Recall  & 0.44816  & 0.46750   & 0.50628    & 0.45818          & 0.48660  & \textbf{0.64852}    & \emph{0.79060}   & 0.59954 \\
\rowcolor[HTML]{EFEFEF}
Precision    & 0.99137     & 0.99799  & 0.99870 &  \textbf{0.99980} &   0.99966  &0.99922    & 0.99352   & 0.98875     \\
F1 Score  & 0.61728   & 0.63673  & 0.67193      & 0.62827    & 0.65460 & \textbf{0.78669} & \emph{0.88052} & 0.74646   \\
\rowcolor[HTML]{EFEFEF}
Median Time to Detect (s) & 33 & \textbf{0} & 1       & 55    & 99\tnote{1} & 4  & n/a & n/a \\
\bottomrule
\end{tabular}
        \begin{tablenotes}
        \small
        \item[1] Baseline 1 detector seemingly never analyzed files until execution. As execution occurred a mean 90s after the trial began, this detector used on average 9s to analyze and make a decision on an executing file. Bold figures indicate the best of the six head-to-head tested tools in total cost, recall, precision, and F1 score. \\
        Underlined figures indicate the network-level detector (network) or supervised learning algorithm (algorithm) outperformed all others. 
        \end{tablenotes}
\end{threeparttable}
\end{table*}

\textbf{Detection Costs---}As with the resource costs, we compute the detection costs from each of the 100K test files, then we scale the average files costs (reported in the bottom half of Table \ref{tab:ave-costs-per-file})  for each tool to the estimate the cost expected in the first year of use. 
Detection costs account for: 
\begin{myitemize}
    \item \textit{Triage} (alerts both false and true positives). For every alert a triage cost is \$35.05 = 0.5 hours at \$70/hour (labor cost) + \$0.05 SIEM indexing fee (based on Splunk pricing). 
    \item \textit{Incident Response (IR)} (only for true positives). For IR we charge \$140 = 2 hours at 70/hour (fully burdened cost).
    \item \textit{Attack Costs} (accounts for true positive and false negative alerts and for the time to detection). Attack cost is an increasing function of time that seeks to estimate the cost to the organization from the malware running. 
\end{myitemize}

We model the per-malware attack cost using a custom S curve. 
Let $t_0$ be the malware download time and $t_e$ the execution time of the malware (both measured/recorded in our experiment). Our goal is to identify an S curve $f(t ; t_0, t_e)$ that has a value of 0 for $t<t_0$; is positive beginning at $t = t_0$ but remains small; starts increasing quickly at execution time, $t = t_e$;  hits its inflection at one minute post-execution, $t = t_e + 60$; and approaches a maximum cost of \$1,000 (horizontal asymptote $M =~ $\$1,000).
To find this curve, note that all cumulative distribution functions (CDFs) with sample space $[0, \infty)$ are S shaped increasing functions that begin at $(0,0)$ and approach horizontal asymptote $y = 1$. 
Given our aforementioned goals, we select the CDF of the $\Gamma$ distribution $f(t)$ so $f'(t) = c\ t^{\alpha -1}\exp{(-\beta t)}$ and shift by $t_0$ (replace $f(t)$ with $f(t-t_0)$).   
Then we solve for $\alpha$ and $\beta$ with three constraints.    
(1) $t_e$ is the first root of $f'''$, forcing the curve to be small and begin to grow only at execution time.
(2) $f''(t_e + 60) = 0 $, forcing the curve to be accruing cost most rapidly a minute after execution. 
The solution gives $\alpha = (1 + 60/t_e)^2 +1, \beta = (\alpha - 1)^2 / 60. $
(3) Multiplying $f$ by 1,000 ensures $\lim_{t\nearrow \infty} f = 1,000$, forcing \$1K to be the max accrued cost.  
See Figure \ref{fig:attack-cost}.
This curve is fit to each trial as the execution time, $t_e$ varies.

If a malware sample is detected before execution (i.e., in time $t \in [t_0,t_e)$), attack cost is simply $f(t; t_0, t_e)$ (near 0) and no triage or IR costs are incurred, as we assume the endpoint client quarantines the file. 
Otherwise, if the malware is detected in time $t \in [t_e, t_e + 60]$ (from execution to a minute after), the attack cost is $f(t; t_0, t_e)$ + Triage + IR.
If the VM closed and the detection technology did not yet alert,  we charge the max cost, $1,000  = \lim_{t\to\infty}f(t; t_0, t_e)$. 
Note that in some cases alert logs were received after the VM closed as technologies that leverage a central appliance could have been processing information from the file.
In this case we charge $C$ + Triage + IR, where $C = \texttt{ave(} f(t_e+60; t_0, t_e),$ 1000 - Triage - IR)$)$.
Note this is higher than a detection cost at $f(t_e)$ but less than the max cost of $1,000$ as the detector, given more time, might have eventually alerted. 
This is designed so that the overall cost (attack + IR + triage costs) for detection after the VM closes the experiment is strictly more than during the VM and strictly less than never detecting.
For each tool, after computing the per-malware cost incurred, the average cost per malware is computed.

If a benignware is detected (false positive), we charge a Triage cost; otherwise (true negative) no cost is incurred. 
For each tool, after computing the per-benignware cost,  the average cost per benignware is computed.

\begin{table*}
\centering
\caption{Machine Learning Versus Signature-Based Detection Statistics}
\label{tab:zero-day-table}
\begin{tabular}{llrccccccrcc} 
\toprule
& & & \multicolumn{6}{c}{\textbf{ML-Based}} && \multicolumn{2}{c}{\textbf{Signature-Based}}  \\
\textbf{Statistic}  & \textbf{Samples} && \textbf{Tool 1} & \textbf{Tool 2} & \textbf{Tool 3} & \textbf{Tool 4} & \textbf{Network} & \textbf{Algorithm} && \textbf{Baseline 1} & \textbf{Baseline 2}     \\
\cmidrule{1-2} \cmidrule{4-9} \cmidrule{11-12}
Recall (\% Malware Detected) & Zero-day PEs    && 45.5\%          & 40.1\%          & 38.1\%          & 27.8\%          & 59.4\%           & 59.7\%          & & 3.8\%               & 4.4\%                   \\
\rowcolor[HTML]{EFEFEF}
Recall (\%
  Malware Detected)      & Public PEs      && 80.7\%          & 87.9\%          & 59.8\%          & 71.0\%          & 90.8\%           & 93.6\%             && 68.2\%              & 76.6\%                  \\
Time to Detect Median (s)  & Zero-day PEs  && 36              & 0               & 0               & 47              & -                & -                  && 97                  & 5                       \\
\rowcolor[HTML]{EFEFEF}
Time to Detect Median (s)         & Public PEs      && 33              & 0               & 1               & 50              & -                & -                  && 99                  & 5                       \\
Malware no other tool detected (\#) & Zero-day PEs       &  & 1               & 0               & 35              & 0               & 86               & 61                 && 0                   & 0                       \\
\rowcolor[HTML]{EFEFEF}
Malware no other tool detected (\#)    & Public  PEs &    & 24              & 16              & 9               & 8               & 54               & 468                && 2                   & 11                      \\
\bottomrule
\end{tabular}
\end{table*}

\textbf{Total Cost---}The total cost is the sum of the initial, detection, and, ongoing resource costs for the first year of use. 
The average resource and detection costs per malware and benignware, as reported in Table \ref{tab:ave-costs-per-file}, are each linearly scaled to 50K files---our estimate for the number of files a host detector sees in its first year---with 1.16\% malware and 98.84\% benignware. 
This overall number of files (50K) was informed by SOC operators who pulled endpoint detector logs from $\smalltilde$30 IPs for a full calendar year. 
The 1.16/98.84\% malware/benignware ratio follows Li et al. \cite{li2017large}. 
These annual estimates are shown in Table II in the lines labeled ``Annual Malware Resource Cost,'' ``Annual Benignware Resource Cost,'' ``Annual Malware Detect Cost,'' and ``Annual Benignware Detect Cost.'' 

Note that the the annual malware and benignware resource costs account only for the resource costs of the detector when a file is present. 
To account for the whole year, we simply scale the ambient resource cost estimate observed for this tool to the  the remaining time in the year. 
This is shown in Table II as  ``Annual Ambient Endpoint Resource Cost.'' 
Resource appliance costs are also included.

Because precision is so close to perfect for these tools (as we shall see), the overall costs scale approximately according to an affine function ($mx + b$, here $x$ is the number of files seen or equivalently time in this model)  with the slope $m$ an increasing function of the observed recall, the detection latency, and the max attack cost. 
Over a long enough time (or after encountering enough files), those tools with the best recall and time to detect will prevail.


\section{AI ATAC 1 Results and Discussion} 
\label{sec:aiatac1-results}
This section provides the results of the endpoint malware experiment. 
We present the summarized and itemized takeaways on the findings in the introduction, Section \ref{sec:edr-takeaways}. 

First note that the initial and resource costs pale in comparison to the detection costs, indicating that this cost model differentiates tools based on their detection statistics---precision, recall, and detection time. 
Precision is near perfect for all tools, which is reflected in the low detection costs incurred from benignware, especially considering that our simulated ratio of files is 98.84\% of the files in the simulation. 
Consequently, recall and time to detect are the two measurements that differentiate these detectors. 

As seen in the bottom of Table \ref{tab:results}, Tools 1--4 (all four ML-based endpoint detection tools) and Baseline 1 (signature-based endpoint detection tool) achieve recall at \smalltilde 45\%, whereas Baseline 2 (the second signature-based endpoint detector) has recall near 65\%. 
This combined with the fact that Baseline 2 has a relatively fast detection time (4s) drives the best results in the cost model, and this can been seen beginning in Table \ref{tab:ave-costs-per-file} where Baseline 2 has the lowest average malware detection cost (\$353/malware), leading to the lowest annual malware cost (\$204K = \$$353 \times$50K files/y $\times$ 1.16 malware/100 files). 
 Tool 3,  which leveraged an on-premises appliance in coordination with the endpoint clients, places second. 
Although this increases the resource costs, it pays off in this cost model as the recall of \smalltilde 50\% and low detection time led to a good cost score. 

Median detection time seems to be a differentiator of these tools, as times vary over two orders of magnitude. 
This has an imprint in the annual resource cost, save noting that the anomalously high resource costs of Tool 3 are due to its use of an on-premises appliance. 

\begin{table*}
\centering 
\begin{threeparttable}
\caption{Recall for each File Type} 
\label{tab:filetype-recall}
\begin{tabular}{lrrrrrrrr}
\toprule
 \textbf{Filetype (\# malware)}&            \textbf{Tool 1} &             \textbf{Tool 2} &             \textbf{Tool 3 }&             \textbf{Tool 4} &         \textbf{Baseline 1} &        \textbf{ Baseline 2} &           \textbf{Network} &         \textbf{Algorithm} \\
\midrule
Compressed (3,252)  & 31.55\% & 0.37\% & 31.98\% & 2.31\% & 5.32\% & 44.59\% & 51.66\% & 33.06\% \\
\rowcolor[HTML]{EFEFEF}
HTML (4,575)  & 0.00\% & 0.00\% & 44.85\% & 8.07\% & 0.00\% & 64.90\% & 73.11\% & 0.00\% \\
Image (986)& 0.00\% & 0.00\% & 42.29\% & 7.00\% & 66.53\% & 5.58\% & 68.76\% & 0.00\% \\
\rowcolor[HTML]{EFEFEF}
JAR (127)& 0.00\% & 0.00\% & 5.51\% & 0.00\% & 0.00\% & 0.00\% & 40.16\% & 0.00\% \\
MS-Office (299) & 0.00\% & 49.16\% & 51.51\% & 47.83\% & 54.85\% & 81.94\% & 71.57\% & 79.26\% \\
\rowcolor[HTML]{EFEFEF}
PDF (3,977) & 0.00\% & 0.00\% & 59.67\% & 33.77\% & 0.00\% & 92.33\% & 96.18\% & 95.35\% \\
PE (26,930) & 79.40\% & 86.15\% & 58.98\% & 69.38\% & 66.97\% & 73.90\% & 89.66\% & 92.36\% \\
\rowcolor[HTML]{EFEFEF}
Source-code (905) & 0.00\% & 0.00\% & 31.38\% & 12.93\% & 48.07\% & 49.94\% & 54.14\% & 0.00\% \\
Text (8,738) & 0.00\% & 0.00\% & 34.25\% & 23.85\% & 59.21\% & 40.59\% & 56.50\% & 0.00\% \\
\rowcolor[HTML]{EFEFEF}
XML (211) & 0.00\% & 7.11\% & 53.08\% & 11.85\% & 0.00\% & 63.51\% & 78.20\% & 0.00\% \\
\bottomrule
\end{tabular}
\end{threeparttable}
\end{table*}

Considering the network dynamic, ML-based detector, we see an enormous increase in recall to \smalltilde 79\% with no sacrifice to precision.  This shows the enormous gain in detection capabilities provided by this tool. 
This network tool was evaluated in previous work \cite{bridges2022beyond} for detection latency, where the authors found ``While seeing data rates of up to 7 GB/s, the [tool] achieved median detection [time] from file delivery until alert of ... 258s = 4.3m.''
Hence, the increase in detection capabilities comes at the cost of both latency and the various drawbacks associated with network-level detection; in particular, the need for decrypted traffic and files, the inability to immediately quarantine files, and detection capabilities varying across protocols. 

The algorithm column in Table II, representing the results of a supervised ML algorithm submitted by a cyber technology vendor, also  has a recall of \smalltilde 60\% and near perfect precision. 
We can report that  latency between submission of a file to the algorithm and receiving the detection results had  a median of 1s. Yet, this is not a direct comparison with any other tools, as our experiment required Tools 1--4 and Baselines 1 and 2 to automatically identify the file in a running VM and then analyze it.

\subsubsection{ML Versus Signature-Based Detection}
\label{sec:ml-v-sig}
Our data set includes \smalltilde 1K zero-day (i.e., never before seen) malware, all of which are PEs. We compare results on these \smalltilde 1K zero-day PEs with the analogous results on the \smalltilde 26K public malware PEs. 
Approximately 23\% of these zero-days were not detected by any tool in this experiment, whereas only 2\% of the public PEs were not detected by any tool. 
Table \ref{tab:zero-day-table} provides many statistics comparing the six ML-based detectors to the two signature-based detectors on the zero-day and public PE files. 
As indicated by the recall on zero-day files, signature-based tools identified \smalltilde 4\%, whereas the recall of the ML tools was in the 28\%--60\% range. 
Our results also confirm that detection latency is not affected by zero-day files. 
Finally, we report an interesting statistic for each tool, the number of malware no other tool detected, showing that the Tool 3, Network, and Algorithm columns have good capability to complement other detectors. 
Our main takeaway from Table \ref{tab:zero-day-table} is that ML-based tools are much better for detecting never-before-seen malware.

A drawback of cost-benefit simulations, in particular ours, is accuracy of input parameters, e.g., we cannot accurately estimate the quantity and cost of malware to be observed in the wild. The consolation is that one can vary these unknowns and observe the effect on the results. 
As configured, the cost model made no accommodation for zero-day vs. public malware, so this benefit to ML-based tools is not reflected in the costs. 
Increasing the maximum attack cost for only the zero-day files penalizes false negatives and late detection of zero-days over that of public files. 
For this experiment, the percent of malware that is zero-day, or equivalently, the number of zero-days expected in a given year, is needed for inclusion of an increased attack cost for zero-day files.  
For example, if we suppose 1\% of malware are zero-days (giving 6 zero-day malware,  572 public malware, and 49,420 benignware per year) and increase the maximum attack cost from \$1K for public malware to  \smalltilde \$35,500, then the ML-based tools will begin to surpass both baseline tools in the cost simulation.

\subsubsection{Results Per File Type} 
We present detection rates per file type for all eight tools on our much larger corpus in Table \ref{tab:filetype-recall}. Interestingly, Tool 1 seems to support only PE and compressed file types, while Tool 3 and the network tool have relatively good recall on nearly all file types. 
In short, this is a large discriminator for modern malware detection technologies as efficacy varies wildly. 
Our evidence supports previous findings \cite{bridges2022beyond} that most excel on PE files and vary widely on other types.

\subsection{Drawbacks and Potential Future Research}
\label{drawbacks}
Drawbacks of the software engineering and design choices are consolidated in Section \ref{sec:technical-drawbacks}. 
Conditioned on file type, the benign-malware ratio is skewed because of our inability to find enough benign files of certain types. 
It is unclear how well our eight tested tools represent the  market space of endpoint malware detectors. 
Notably, there was diversity in the types of detectors used---two signature-based detectors, multiple detectors that claimed to \textit{only} use supervised ML (as opposed to those that leverage both a library of signatures and ML), one dynamic (albeit network-level detector), and one that leveraged a central server to assist decision-making by the per-host clients. 
No effort was made to ensure the malware would execute correctly during the dynamic phase of our test.
No internet connection was allowed in this competition, eliminating detectors that leverage a cloud connection and possibly affecting malware executions. 
No effort was made to validate the labels on our samples from expected malicious/benign sources. 

Although each input cost to the model is believable, the final numbers seem high, and there are several possible reasons for inflation of the total cost, most notably, attack cost is notoriously hard to estimate. 
For mitigation, we vary the maximum attack cost parameter  and observe its effect. As tools have roughly 50\% recall on average, the consequence is they incur the max attack cost (\$1K) for about 250 samples (50K files/year $\times$ 1.16/100 malware $\times$ 50\%), which totals  $\smalltilde$ \$250K, plus, the small detection costs for the remaining 50\% malware. 
Extrapolation also causes errors---this cost estimate, although perhaps accurate for the first host in the network affected by malware, likely would not scale to every host in the whole network; rather, once-seen malware would automatically be blocked using modern endpoint detection configurations. 
This means the cost model as configured is a worst-case scenario in that it assumes novel malware per host, which is unlikely. 
Perhaps the malware in this experiment is harder to detect than in the real world, and the percent of zero-days is 1K/50K = 2\%, possibly higher than in the wild. 
A large-scale study to understand rates of zero-day attacks would be useful.  
In a real SOC, the cumulative IR and triage costs are limited by logistics, namely the time available for triage and IR by operators; yet, our cost model does not take this into account.      
Although the IR and triage rates are (we believe) good estimates for a single investigation, they might not be reasonable for influxes (i.e., a flurry) of alerts. In such  cases, it is likely many alerts would simply go unfilled without the operators needed to investigate all of them. Further, it is possible that many similar alerts would be handled in bulk, leveraging automation capabilities of modern endpoint tools (or SOAR tools).
In short, the cost model linearly adds a fixed amount for each alert/TP, but this might not be realistic in that it overestimates costs in some scenarios. 
Operations or perhaps qualitative research of SOCs is needed to understand and model the nonlinearity of their processes. 
All together, these  cost-model drawbacks are particular instances of the overarching problems with cost-benefit analyses---untenable assumptions are needed to model real-world processes for which we have little data. 


\small
\bibliographystyle{IEEEtran}
\bibliography{refs}
\end{document}